\def\E{\end{document}}
\documentclass[11pt]{article}
\usepackage{amssymb,amsmath}

\begin{document}
\title{\bf
The role of potential, Morawetz estimate and spacetime bound for quasilinear Schr\"{o}dinger equations}
  \author{Xianfa Song{\thanks{E-mail: songxianfa@tju.edu.cn(X.F. Song)
 }} \\
\small Department of Mathematics, School of Mathematics, Tianjin University,\\
\small Tianjin, 300072, P. R. China
}

\maketitle
\date{}

\newtheorem{theorem}{Theorem}[section]
\newtheorem{definition}{Definition}[section]
\newtheorem{lemma}{Lemma}[section]
\newtheorem{proposition}{Proposition}[section]
\newtheorem{corollary}{Corollary}[section]
\newtheorem{remark}{Remark}[section]
\renewcommand{\theequation}{\thesection.\arabic{equation}}
\catcode`@=11 \@addtoreset{equation}{section} \catcode`@=12

\begin{abstract}

In this paper, we deal with the following Cauchy problem
\begin{equation*}
\left\{
\begin{array}{lll}
iu_t=\Delta u+2uh'(|u|^2)\Delta h(|u|^2)+V(x)u,\ x\in \mathbb{R}^N,\ t>0\\
u(x,0)=u_0(x),\quad x\in \mathbb{R}^N.
\end{array}\right.
\end{equation*}
Here $h(s)$ and $V(x)$ are some real functions. We take the potential $V(x)\in L^q(\mathbb{R}^N)+L^{\infty}(\mathbb{R}^N)$ as criterion of the blowup and global existence of the solution to (\ref{1}). In some cases, we can classify it in the following sense: If $V(x)\in S(I)$, then the solution of (\ref{1}) is always global existence for any $u_0$ satisfying $0<E(u_0)<+\infty$; If $V(x)\in S(II)$, then the solution of (\ref{1}) may blow up for some initial data $u_0$. Here $$ S(I)=\cup_{q>q_c}[L^q(\mathbb{R}^N)+L^{\infty}(\mathbb{R}^N)],\quad S(II)=\left\{\cup_{q<q_c}[L^q(\mathbb{R}^N)+L^{\infty}(\mathbb{R}^N)]\right\}\setminus S(I).$$

Under certain assumptions, we also establish Morawetz estimates and spacetime bounds for the global solution, for example,
\begin{align*}
 &\int_0^{+\infty}\int_{\mathbb{R}^N}\frac{[|\nabla h(|u|^2)|^2+|V(x)||u|^2]}{(|x|+t)^{\lambda}}dxdt\leq C,\\
&\|u\|_{L^{\bar{q}}_t(\mathbb{R})L^{\bar{r}}_x(\mathbb{R}^N)}=\left(\int_0^{+\infty}\left(\int_{\mathbb{R}^N}|u|^{\bar{r}}dx\right)^{\frac{\bar{q}}{\bar{r}}}dt\right)^{\frac{1}{\bar{q}}}\leq C.
\end{align*}



\end{abstract}

\section{Introduction}
\qquad In this paper, we consider the following Cauchy problem:
\begin{equation}
\label{1} \left\{
\begin{array}{lll}
iu_t=\Delta u+2uh'(|u|^2)\Delta h(|u|^2)+V(x)u, \ x\in \mathbb{R}^N, \ t>0\\
u(x,0)=u_0(x),\quad x\in \mathbb{R}^N.
\end{array}\right.
\end{equation}
Here $N\geq 3$, $h(s)$ and $V(x)$ are some real functions, and $h(s)\geq 0$ for $s\geq 0$. (\ref{1}) can be applied to illustrate many physical phenomena, such as the superfluid film  equation in plasma  physics if $h(s)=s$ and the self-channelling of a high-power ultra short laser in matter if $h(s)=\sqrt{1+s}$. It also appears in condensed matter theory and in dissipative quantum mechanics, see \cite{Bass, Borovskii,  Bouard, Goldman, Litvak, Makhankov, Ritchie}. The local well-posedness of quasilinear Sch\"{o}dinger equation has been studied by many authors, see \cite{Colin1,Kenig, Poppenberg1} and the references therein. An interesting topic on (\ref{1}) is the global existence and blowup phenomena of the solution. The definition of the global existence and blowup in finite time for the solution of (\ref{1}) is given as follows.

{\bf Definition 1.} {\it Let $u(x,t)$ be the solution of (\ref{1}). We say that $u(x,t)$ will blow up in finite time if there exists a time $0<T<+\infty$ such that
\begin{align}
\lim_{t\rightarrow T^-} \int_{\mathbb{R}^N}[|\nabla u(x,t)|^2+|\nabla h(|u(x,t)|^2)|^2)]dx=+\infty.
\end{align}
Otherwise, we say that $u(x,t)$ exists globally if the maximal existence interval for $t$ is $[0, +\infty)$.
 }

About the topic on the global existence and blowup phenomena of semilinear Schr\"{o}dinger equation, the following Cauchy problem
\begin{equation}
\label{2} \left\{
\begin{array}{lll}
iu_t=\Delta u+F(|u|^2)u,\quad x\in \mathbb{R}^N, \ t>0\\
u(x,0)=u_0(x),\quad x\in \mathbb{R}^N
\end{array}\right.
\end{equation}
 was considered by Glassey in his famous paper \cite{Glassey}. $sF(s)\geq c_N G(s)$ for some constant $c_N>1+\frac{2}{N}$ and all $s\geq 0$ is the key condition on the blowup of the solution to (\ref{2}). The related results on semilinear Schr\"{o}dinger equation also can be found in the book \cite{Cazenave} and the references therein. However, there are very few conferences about the conditions on the global existence and blowup of the solution to a quasilinear Schr\"{o}dinger equation, we can refer to \cite{Bouard, Guo, SW2}.

This paper is a parallel one to \cite{SW2}. Very recently, we studied the following problem
\begin{equation}
\label{1'} \left\{
\begin{array}{lll}
iu_t=\Delta u+2uh'(|u|^2)\Delta h(|u|^2)+(W*|u|^2)u, \quad x\in \mathbb{R}^N, \ t>0\\
u(x,0)=u_0(x),\quad x\in \mathbb{R}^N
\end{array}\right.
\end{equation}
in \cite{SW2}. We took $W(x)$ as the criterion and focused on how the potential $W(x)$ in Hartree type nonlinear term takes effect on the properties for the solution. Naturally, we are interested in the conditions on $V(x)$ which can make the solution of (\ref{1}) blow up in finite time or exist globally--- that is the first motivation of this paper.

Before we state our main results, we define the mass and energy below. Their conservation laws will be proved in Section 2.

Mass: $$ m(u)=\left(\int_{\mathbb{R}^N}|u(x,t)|^2dx\right)^{\frac{1}{2}};$$

Energy: $$E(u)=\frac{1}{2}\int_{\mathbb{R}^N}[|\nabla u|^2+|\nabla h(|u|^2)|^2]dx-\frac{1}{2}\int_{\mathbb{R}^N}V(x)|u|^2dx.$$

Our first result is about the sufficient conditions on the blowup in finite time for the solution of (\ref{1}).

{\bf Theorem 1.} {\it Let $u(x,t)$ be the solution of (\ref{1}) with $u_0\in X$, $xu_0\in L^2(\mathbb{R}^N)$, $-\infty<E(u_0)\leq 0$ and $\Im \int_{\mathbb{R}^N}\bar{u}_0(x\cdot \nabla u_0)dx>0$.
Here
\begin{align}
 X=\{w\in H^1(\mathbb{R}^N),\quad \int_{\mathbb{R}^N}|\nabla h(|w|^2)|^2dx<+\infty\}.\label{kongjianziji}
\end{align}
 Assume that there exists constant $k$ such that $sh''(s)\leq kh'(s)$ if $h'(s)\geq 0$ or  $sh''(s)\geq kh'(s)$ if $h'(s)\leq 0$. If $V(x)=V_1(x)+V_2(x)\in L^q(\mathbb{R}^N)+L^{\infty}(\mathbb{R}^N)$ for some $q>1$ and $$(C1)\qquad \qquad \quad \quad[\max((2k+1)N,0)+2]V+(x\cdot \nabla V)\leq 0,\qquad \qquad \qquad \qquad$$ then there exists a finite time $T$ such that
$$
\lim_{t\rightarrow T^-} \int_{\mathbb{R}^N}[|\nabla u(x,t)|^2+|\nabla h(|u|^2)(x,t)|^2]dx=+\infty.
$$
}

Our second result will establish the sufficient conditions on the global existence of the solution to (\ref{1}).

{\bf Theorem 2.} {\it  Let $u(x,t)$ be the solution of (\ref{1}) with $u_0\in X$.

(1). If $V(x)=V_1(x)+V_2(x)\in L^q(\mathbb{R}^N)+L^{\infty}(\mathbb{R}^N)$ for some $q>1$ and $V(x)\leq 0$ for all $x\in \mathbb{R}^N$, then $u$ is global existence for any initial $u_0\in X$ satisfying $0<E(u_0)<+\infty$.

(2). Assume that $V(x)\geq 0$ for all $x\in\mathbb{R}^N$ or changes sign, and there exist $a>0$ in the sense of infimum and $\alpha>0$ in the sense of supremum such that $\max(s^{\frac{1}{2}}, s^{\alpha})\leq a[h(s)+s^{\frac{1}{2}}]$ for $s\geq 1$, $$
(C2)\quad V(x)=V_1(x)+V_2(x)\in L^q(\mathbb{R}^N)+L^{\infty}(\mathbb{R}^N),\ q>1,\ q\geq \frac{2^*}{\max(2\alpha, 1) 2^*-2}. $$

We have

Case (i) $0<\alpha\leq\frac{1}{2}$, $q>\frac{N}{2}$, then the solution is global existence for any initial $u_0\in X$ satisfying $0<E(u_0)<+\infty$;

or

Case (ii) $0<\alpha\leq \frac{1}{2}$, $q=\frac{N}{2}$, if $$(C_s)^{\frac{2}{2^*}}\left(\int_{\mathbb{R}^N}|V_1(x)|^{\frac{N}{2}}dx\right)^{\frac{2}{N}}<1,$$ then the solution is global existence for any initial $u_0\in X$ satisfying $0<E(u_0)<+\infty$. Here $C_s$ be the best constant in the Sobolev's inequality
\begin{align}
\int_{\mathbb{R}^N}w^{2^*}dx\leq C_s\left(\int_{\mathbb{R}^N}|\nabla w|^2dx\right)^{\frac{2^*}{2}}\quad {\rm for \ any}\quad w\in H^1(\mathbb{R^N}).\label{zjcs}
\end{align}

or

Case (iii) $\frac{1}{2}<\alpha<\frac{N-1}{N}$, $q>\frac{2^*}{\max(2\alpha, 1)2^*-2}$, then the solution is global existence for any initial $u_0\in X$ satisfying $0<E(u_0)<+\infty$.

or

Case (iv) $\alpha\geq\frac{N-1}{N}$, $q\geq \frac{\alpha\cdot 2^*}{\alpha\cdot2^*-1}$, then the solution is global existence for any initial $u_0\in X$ satisfying $0<E(u_0)<+\infty$.

}

{\bf Remark 1.1.}  1. If $h(|u|^2)\equiv 0$, (\ref{1}) becomes the classic linear Schr\"{o}dinger with potential, we can take $a=1$ and $\alpha=\frac{1}{2}$, the solution of (\ref{1}) is global existence if $V(x)\in L^q(\mathbb{R}^N)+L^{\infty}(\mathbb{R}^N)$, $q>\frac{N}{2}$. On the other hand, we can take $k=-\frac{1}{2}$ in Theorem 1. If $2V+(x\cdot \nabla V)\leq 0$ and $E(u_0)<0$, the solution of (\ref{1}) will blow up in finite time, which meets with the classic result in Chapter 6 of \cite{Cazenave}. Especially, if $V(x)=\frac{1}{|x|^m}(x\neq \mathbf{0})$, $m\geq 2$, we can call it as ``{\bf the blowup phenomenon deduced by the singular potential in the classic linear Schr\"{o}dinger equation}".

2. If $h(s)=s^{\alpha}$, we can take $k=\alpha-1$ in Theorem 1 and condition (C1) becomes $[\max((2\alpha-1)N,0)+2]V+(x\cdot \nabla V)\leq 0$. If $0<\alpha<\frac{N-1}{N}$, then the critical exponent is $q_c=\frac{2^*}{\max(2\alpha, 1)2^*-2}$ in the following sense: There exists initial $u_0$ such that the solution will blow up in finite time when
$q<q_c$ by Theorem 1, while the solution is global existence for any initial $u_0\in X$ when $q>q_c$ by Theorem 2.

In another word, the function $V(x)\in L^q(\mathbb{R}^N)+L^{\infty}(\mathbb{R}^N)$ can be classified and taken as criterion of the blowup and global existence of the solution to (\ref{1}) in the following sense: If $V(x)\in S(I)$, then the solution of (\ref{1}) is always global existence for any initial data $u_0$; If $V(x)\in S(II)$, then the solution of (\ref{1}) may blow up for some initial data $u_0$. Here $$ S(I)=\cup_{q>q_c}[L^q(\mathbb{R}^N)+L^{\infty}(\mathbb{R}^N)],\quad S(II)=\left\{\cup_{q<q_c}[L^q(\mathbb{R}^N)+L^{\infty}(\mathbb{R}^N)]\right\}\setminus S(I).$$
Regrettably, if $\frac{1}{2}<\alpha<\frac{N-1}{N}$ and $V(x)\in[L^{q_c}(\mathbb{R}^N)+L^{\infty}(\mathbb{R}^N)]$, we cannot make sure whether the solution is global existence or not due to the technical difficulty. We left it as an open problem.

It is very interesting to consider the asymptotic behaviors and the spacetime bounds for the solution of (\ref{1}). Basing on the pseudo-conformal conservation laws, we also obtain the results on the asymptotic behavior and spacetime bounds for the global solution and lower bound for the blowup rate of the blowup solution, which is the second motivation of this paper. For example, under certain assumptions, we can get
$$
\int_0^{+\infty}\int_{\mathbb{R}^N}\frac{[|\nabla h(|u|^2)|^2+|V(x)||u|^2]}{(|x|+t)^{\mu}}dxdt\leq C,
$$
and
$$
\|u\|_{L^{\bar{q}}_t(\mathbb{R})L^{\bar{r}}_x(\mathbb{R}^N)}=\left(\int_0^{+\infty}\left(\int_{\mathbb{R}^N}|u|^{\bar{r}}dx\right)^{\frac{\bar{q}}{\bar{r}}}dt\right)^{\frac{1}{\bar{q}}}\leq C.
$$
To control the length of introduction, we would like to state these results in Section 4.

The organization of this paper is as follows. In Section 2, we will prove the mass and energy conservation laws, give the proof of Theorem 1.
In Section 3, we will prove Theorem 2. In Section 4, we will establish a pseudoconformal conservation law and consider the asymptotic behaviors and spacetime bounds for the solution.

\section{Preliminaries and the proof of Theorem 1}
\qquad In the sequels, we use $C$, $C'$, and so on, to denote some constants, the values may vary line to line.

First, we will give a lemma as follows.

{\bf Lemma 2.1.} {\it Assume that $u$ is the solution of (\ref{1}). Then in the time interval $[0,t]$ when it exists, $u$ satisfies

(i) Mass conversation: $$ m(u)=\left(\int_{\mathbb{R}^N}|u(x,t)|^2dx\right)^{\frac{1}{2}}=\left(\int_{\mathbb{R}^N}|u_0(x)|^2dx\right)^{\frac{1}{2}}=m(u_0);$$

(ii) Energy conversation: $$E(u)=\frac{1}{2}\int_{\mathbb{R}^N}[|\nabla u|^2+|\nabla h(|u|^2)|^2]dx-\frac{1}{2}\int_{\mathbb{R}^N}V(x)|u|^2dx=E(u_0);$$

(iii) $$\frac{d}{dt} \int_{\mathbb{R}^N}|x|^2|u|^2dx=-4\Im \int_{\mathbb{R}^N} \bar{u}(x\cdot \nabla u)dx;$$

(iv) \begin{align}
&\quad \frac{d}{dt} \Im \int_{\mathbb{R}^N} \bar{u}(x\cdot \nabla u)dx=-2\int_{\mathbb{R}^N}|\nabla u|^2dx-(N+2)\int_{\mathbb{R}^N}|\nabla h(|u|^2)|^2dx\\
&\qquad -8N\int_{\mathbb{R}^N}h''(|u|^2)h'(|u|^2)|u|^4|\nabla u|^2dx-\int_{\mathbb{R}^N}(x\cdot \nabla V)|u|^2dx.\label{10131}
\end{align}

}

{\bf Proof:} (i) Multiplying (\ref{1}) by $2\bar{u}$, taking the imaginary part of the result, we get
\begin{align}
\frac{\partial }{\partial t}|u|^2=\Im(2\bar{u}\Delta u) =\nabla \cdot (2\Im \bar{u}\nabla u).\label{10121}
\end{align}
Integrating it over $\mathbb{R}^N\times [0,t]$, we have
$$ \int_{\mathbb{R}^N}|u|^2dx=\int_{\mathbb{R}^N}|u_0|^2dx$$
and take arithmetic square root on both sides to have mass conservation law.

(ii)  Multiplying (\ref{1}) by $2\bar{u}_t$, taking the real part of the result, then integrating it over $\mathbb{R}^N\times [0,t]$, we obtain
\begin{align*}
&\quad\int_{\mathbb{R}^N}[|\nabla u|^2+|\nabla h(|u|^2)|^2]dx-\int_{\mathbb{R}^N}V(x)|u|^2dx\\
&=\int_{\mathbb{R}^N}[|\nabla u_0|^2+|\nabla h(|u_0|^2)|^2]dx-\int_{\mathbb{R}^N}V(x)|u_0|^2dx
\end{align*}
and multiply $\frac{1}{2}$ on both sides to reach energy conservation law.

(iii) Multiplying (\ref{10121}) by $|x|^2$ and integrating it over $\mathbb{R}^N$, we get
\begin{align*}
\frac{d}{dt}\int_{\mathbb{R}^N}|x|^2|u|^2dx&=\int_{\mathbb{R}^N}|x|^2\nabla \cdot(2\Im (\bar{u}\nabla u))dx
=-4\Im \int_{\mathbb{R}^N}\bar{u}(x\cdot \nabla u)dx.
\end{align*}

(iv) Let  $a(x,t)=\Re u(x,t)$ and $b(x,t)=\Im u(x,t)$. Then $u(x,t)=a(x,t)+ib(x,t)$,
$$
\frac{d}{dt}\Im \bar{u}(x\cdot \nabla u)=\sum_{k=1}^N[x_k(b_t)_{x_k}a-x_k(a_t)_{x_k}b]+\sum_{k=1}^N(x_kb_{x_k}a_t-x_ka_{x_k}b_t).
$$
And
\begin{align*}
 \frac{d}{dt}\Im \int_{\mathbb{R}^N}\bar{u}(x\cdot \nabla u)dx&=\int_{\mathbb{R}^N} \sum_{k=1}^N[x_k(b_t)_{x_k}a-x_k(a_t)_{x_k}b]dx
+\int_{\mathbb{R}^N} \sum_{k=1}^N (x_ka_{x_k}\Delta a+x_kb_{x_k}\Delta b)dx\nonumber\\
&\quad+\frac{1}{2}\int_{\mathbb{R}^N}\sum_{k=1}^N x_k(|u|^2)_{x_k}[2h'(|u|^2)\Delta h(|u|^2)+V(x)]dx\nonumber\\
&=N\int_{\mathbb{R}^N}(a_tb-ab_t)dx+\int_{\mathbb{R}^N}\sum_{k=1}^N(x_kb_{x_k}a_t-x_ka_{x_k}b_t)dx\nonumber\\
&\quad+\frac{N-2}{2}\int_{\mathbb{R}^N}|\nabla u|^2dx+\frac{N-2}{2}\int_{\mathbb{R}^N}|\nabla h(|u|^2)|^2dx\nonumber\\
&\quad-\frac{1}{2}\int_{\mathbb{R}^N}[NV+(x\cdot \nabla V)]|u|^2dx\nonumber\\
&=N\int_{\mathbb{R}^N}\left([a\Delta a+b\Delta b]+2|u|^2h'(|u|^2)\Delta h(|u|^2)+V(x)|u|^2\right)dx\nonumber\\
&\quad +(N-2)\int_{\mathbb{R}^N}|\nabla u|^2dx+(N-2)\int_{\mathbb{R}^N}|\nabla h(|u|^2)|^2dx\nonumber\\
&\quad-\int_{\mathbb{R}^N}[NV+(x\cdot \nabla V)]|u|^2dx\nonumber\\
&=-2\int_{\mathbb{R}^N}|\nabla u|^2dx-(N+2)\int_{\mathbb{R}^N}|\nabla h(|u|^2)|^2dx\nonumber\\
&\quad-8N\int_{\mathbb{R}^N}h'(|u|^2)h''(|u|^2)|u|^4|\nabla u|^2dx-\int_{\mathbb{R}^N}(x\cdot \nabla V)|u|^2dx.
\end{align*}
Lemma 2.1 is proved.\hfill $\Box$

 Now we deal with the sufficient conditions on blowup in finite time for the solution by using the results of Lemma 2.1.

{\bf Proof of Theorem 1:} Wherever $u$ exists, let
$$
y(t)=\Im \int_{\mathbb{R}^N}\bar{u}(x\cdot \nabla u)dx.
$$

We discuss it in two cases:

Case 1. $h(s)\equiv 0$ or $h(s)\neq 0$ and $k\leq -\frac{1}{2}$. We have
\begin{align}
\dot{y}(t)&=-2\int_{\mathbb{R}^N}|\nabla u|^2dx-(N+2)\int_{\mathbb{R}^N}|\nabla h(|u|^2)|^2dx-8N\int_{\mathbb{R}^N}h''(|u|^2)h'(|u|^2)|u|^4|\nabla u|^2dx\nonumber\\
&\quad -\int_{\mathbb{R}^N}(x\cdot \nabla V)|u|^2dx\nonumber\\
&\geq -2\int_{\mathbb{R}^N}|\nabla u|^2dx-(N+2+2kN)\int_{\mathbb{R}^N}|\nabla h(|u|^2)|^2dx-\int_{\mathbb{R}^N}(x\cdot \nabla V)|u|^2dx\nonumber\\
&=-4E(u_0)-(2k+1)N\int_{\mathbb{R}^N}|\nabla h(|u|^2)|^2dx-\int_{\mathbb{R}^N}[2V+(x\cdot \nabla V)]|u|^2dx\nonumber\\
&\geq 0,\label{10141}
\end{align}
which means that $y(t)\geq y(0)>0$ for $t>0$.

Case 2. $h(s)\neq 0$ and $k>-\frac{1}{2}$. We have
\begin{align}
\dot{y}(t)&\geq -2\int_{\mathbb{R}^N}|\nabla u|^2dx-(N+2+2kN)\int_{\mathbb{R}^N}|\nabla h(|u|^2)|^2dx-\int_{\mathbb{R}^N}(x\cdot \nabla V)|u|^2dx\nonumber\\
&=(2k+1)N\int_{\mathbb{R}^N}|\nabla u|^2dx-2[(2k+1)N+2]E(u_0)\nonumber\\
&\quad-\int_{\mathbb{R}^N}[((2k+1)N+2)V+(x\cdot \nabla V)]|u|^2dx\nonumber\\
&\geq 0,\label{10141}
\end{align}
which also means that $y(t)\geq y(0)>0$ for $t>0$.

Setting
$$
J(t)=\int_{\mathbb{R}^N}|x|^2|u|^2dx,
$$
we have $J'(t)=-4y(t)<-4y(0)<0$. Then
$$0\leq J(t)=J(0)+\int_0^tJ'(\tau)d\tau<J(0)-4y(0)t,$$
which implies that the maximum existence interval of time for $u$ is finite, and $u$ will blow up before $\frac{J(0)}{4y(0)}$.\hfill $\Box$

{\bf Remark 2.1.} If $V(x)$ is a nontrivial radially symmetric positive function, we can deduce $$V(r)\geq \frac{V(1)}{r^{[\max((2k+1)N,0)+2]}}$$ for $0<r\leq 1$ by $[\max((2k+1)N,0)+2]V(r)+rV'(r)\leq 0$. However, $$\frac{V(1)}{r^{[\max((2k+1)N,0)+2]}}\notin L^{q}(\mathbb{R}^N)+L^{\infty}(\mathbb{R}^N)$$ for any  $q>\frac{N}{\max[(2k+1)N,0]+2}$, there exists $\tilde{q}<\frac{N}{\max[(2k+1)N,0]+2}$ such that
$$\frac{1}{r^{[\max((2k+1)N,0)+2]}}\in L^{\tilde{q}}(\mathbb{R}^N)+L^{\infty}(\mathbb{R}^N).$$

If $h(s)=s^{\alpha}$, we can take $k=\alpha-1$ in Theorem 1,  then $$\max((2k+1)N,0)+2=\max((2\alpha-1)N,0)+2$$ and $$\frac{N}{\max[(2k+1)N,0]+2}=\frac{N}{\max((2\alpha-1)N,0)+2}=\frac{2^*}{\max(2\alpha, 1)2^*-2}.$$

\section{The proof of Theorem 2}

\qquad In this section, we will prove Theorem 2 and establish the sufficient conditions on the global existence of the solution to (\ref{1}).

{\bf Proof of Theorem 2:}

 {\bf Case (1).} $V(x)\leq 0$ for $x\in \mathbb{R}^N$ and $0<E(u_0)<+\infty$. The global existence of the solution is a direct result of the energy conversation law of Lemma 2.1(ii) because
$$
\int_{\mathbb{R}^N}|\nabla u|^2dx+\int_{\mathbb{R}^N}|\nabla h(|u|^2)|^2dx\nonumber\\
+\int_{\mathbb{R}^N}|V(x)||u|^2dx=2E(u_0)<+\infty,
$$
which implies that $\int_{\mathbb{R}^N}|\nabla u|^2dx+\int_{\mathbb{R}^N}|\nabla h(|u|^2)|^2dx+\int_{\mathbb{R}^N}|V(x)||u|^2dx$ is uniformly bounded for all $t>0$.

{\bf Case (2).} There exist constants $a>0$ in the sense of infimum and $\alpha>0$ in the sense of supremum such that $a[h(s)+s^{\frac{1}{2}}]\geq \max(s^{\frac{1}{2}}, s^{\alpha})$ for $s\geq 1$, $V(x)\geq 0$ for all $x\in\mathbb{R}^N$ or changes sign,  $V\in L^q(\mathbb{R}^N)+L^{\infty}(\mathbb{R}^N)$, $q>1$, $q\geq \frac{2^*}{\max(2\alpha, 1) 2^*-2}$.

Suppose that $V(x)=V_1(x)+V_2(x)$, where $V_1(x)\in L^q(\mathbb{R}^N)$ and $V_2(x)\in L^{\infty}(\mathbb{R}^N)$.

Note that $\max(2\alpha,1)=1$ if $\alpha\leq \frac{1}{2}$, $\max(2\alpha,1)=2\alpha$ if $\alpha>\frac{1}{2}$, while $\frac{2^*}{\max(2\alpha, 1) 2^*-2}>1$
if $0<\alpha<\frac{2^*+2}{22^*}=\frac{N-1}{N}$, $\frac{2^*}{\max(2\alpha, 1) 2^*-2}\leq 1$
if $\alpha\geq \frac{2^*+2}{22^*}=\frac{N-1}{N}$. That is, $\alpha=\frac{1}{2}$ and $\alpha=\frac{N-1}{N}$ are two special points.

We discuss it in different subcases below.

{\bf Subcase (i).} $0<\alpha\leq \frac{1}{2}$, $q>\frac{N}{2}$. Then $\frac{2q}{q-1}<2^*$.
\begin{align*}
&\quad\int_{\mathbb{R}^N}|V_1(x)||u|^2dx
\leq\left(\int_{\mathbb{R}^N}|V_1(x)|^qdx\right)^{\frac{1}{q}}\left(\int_{\mathbb{R}^N}|u|^{\frac{2q}{q-1}}dx\right)^{\frac{q-1}{q}}\\
&\leq \left(\int_{\mathbb{R}^N}|V_1(x)|^qdx\right)^{\frac{1}{q}}\left(\int_{|u|\leq 1}|u|^{\frac{2q}{q-1}}dx\right)^{\frac{q-1}{q}} +\left(\int_{\mathbb{R}^N}|V_1(x)|^qdx\right)^{\frac{1}{q}}\left(\int_{|u|>1}|u|^{\frac{2q}{q-1}}dx\right)^{\frac{q-1}{q}}\nonumber\\
&\leq C_V\left(\int_{|u|\leq 1}|u|^2dx\right)^{\frac{q-1}{q}}+C_V\left(\int_{|u|>1}|u|^2dx\right)^{\frac{[q(2^*-2)-2^*]}{q(2^*-2)}}\left(\int_{|u|>1}|u|^{2^*}dx\right)^{\frac{2}{q(2^*-2)}}.
\end{align*}
By mass and energy conservation laws, using H\"{o}lder inequality, Young inequality, we obtain
\begin{align}
&\int_{\mathbb{R}^N}|\nabla u|^2dx+\int_{\mathbb{R}^N}|\nabla h(|u|^2)|^2dx=2E(u_0)+\int_{\mathbb{R}^N}V(x)|u|^2dx\nonumber\\
&=2E(u_0)+\int_{\mathbb{R}^N}V_1(x)|u|^2dx+\int_{\mathbb{R}^N}V_2(x)|u|^2dx\nonumber\\
&\leq C+\int_{\mathbb{R}^N}|V_1(x)||u|^2dx+\|V_2(x)\|_{L^{\infty}}\int_{\mathbb{R}^N}|u_0|^2dx\nonumber\\
&\leq C'+( C_s)^{\frac{2}{q(2^*-2)}}C_V\left(\int_{\mathbb{R}^N}|u_0|^2dx\right)^{\frac{[q(2^*-2)-2^*]}{q(2^*-2)}}\left(\int_{\mathbb{R}^N}|\nabla u|^2dx\right)^{\frac{2^*}{q(2^*-2)}}\nonumber\\
&\leq C'+\frac{1}{2}\int_{\mathbb{R}^N}|\nabla u|^2dx,\label{1130w3}
\end{align}
which means that
$$
\int_{\mathbb{R}^N}|\nabla u|^2dx+\int_{\mathbb{R}^N}|\nabla h(|u|^2)|^2dx\leq C.
$$

{\bf Subcase (ii).} $0<\alpha\leq \frac{1}{2}$ and $q=\frac{N}{2}$.

Similar to (\ref{1130w3}), we get

\begin{align}
&\quad \int_{\mathbb{R}^N}|\nabla u|^2dx+\int_{\mathbb{R}^N}|\nabla h(|u|^2)|^2dx\nonumber\\
&\leq C'+\left(\int_{\mathbb{R}^N}|V_1(x)|^qdx\right)^{\frac{1}{q}}\left(\int_{\mathbb{R}^N}|u|^{\frac{2q}{q-1}}dx\right)^{\frac{q-1}{q}}\nonumber\\
&=C'+\left(\int_{\mathbb{R}^N}|V_1(x)|^{\frac{N}{2}}dx\right)^{\frac{2}{N}}\left(\int_{\mathbb{R}^N}|u|^{2^*}dx\right)^{\frac{2}{2^*}}\nonumber\\
&\leq C'+(C_s)^{\frac{2}{2^*}}\left(\int_{\mathbb{R}^N}|V_1(x)|^{\frac{N}{2}}dx\right)^{\frac{2}{N}}\int_{\mathbb{R}^N}|\nabla u|^2dx.\label{3141}
\end{align}
If
$$
(C_s)^{\frac{2}{2^*}}\left(\int_{\mathbb{R}^N}|V_1(x)|^{\frac{N}{2}}dx\right)^{\frac{2}{N}}:=C(V,N)<1,
$$
then
$$
\left(\int_{\mathbb{R}^N}|\nabla u|^2dx+\int_{\mathbb{R}^N}|\nabla h(|u|^2)|^2dx\right)\leq \frac{C'}{[1-C(V,N)]}.
$$

{\bf Subcase (iii).} $\frac{1}{2}<\alpha<\frac{2*+2}{22^*}=\frac{N-1}{N}$, $q>\frac{2^*}{2\alpha\cdot2^*-2}$. Then
$\frac{2^*}{q(2\alpha\cdot2^*-2)}<1$.

Denote
\begin{align}
&\theta=\frac{[2-(2\alpha-1)2^*]}{2^*},\quad K=\frac{2q(2\alpha\cdot2^*-2)-2[2-(2\alpha-1)2^*]}{[q(2\alpha\cdot2^*-2)-2^*]},\label{12101}\\
&\frac{1}{\tau_1}=\frac{2^*}{q(2\alpha\cdot2^*-2)},\quad \frac{1}{\tau_2}=\frac{q(2\alpha\cdot2^*-2)-2^*}{q(2\alpha\cdot2^*-2)},\label{12102}\\
&\frac{1}{m_1}=\frac{2\alpha\cdot2^*-K}{2\alpha\cdot2^*-2},\quad \frac{1}{m_2}=\frac{K-2}{2\alpha\cdot2^*-2}.
\end{align}
 Using H\"{o}lder inequality, we get
\begin{align*}
&\quad\int_{\mathbb{R}^N}V(x)|u|^2dx=\int_{\mathbb{R}^N}V_1(x)|u|^2dx+\int_{\mathbb{R}^N}V_2(x)|u|^2dx\nonumber\\
&\leq \int_{\mathbb{R}^N}|V_1(x)||u|^2dx+\|V_2(x)\|_{L^{\infty}}\int_{\mathbb{R}^N}|u_0|^2dx\nonumber\\
&\leq C+\left(\int_{\mathbb{R}^N}|V_1|^{\tau_1}|u|^{2\theta}dx\right)^{\frac{1}{\tau_1}}
\left(\int_{\mathbb{R}^N}|u|^{K}dx\right)^{\frac{1}{\tau_2}}\nonumber\\
&\leq C+\left(\int_{\mathbb{R}^N}|V_1|^qdx\right)^{\frac{1}{q}}\left(\int_{\mathbb{R}^N}|u|^2dx\right)^{\frac{\theta}{\tau_1}}
\left(\int_{\mathbb{R}^N}|u|^2dx\right)^{\frac{1}{\tau_2m_1}}\left(\int_{\mathbb{R}^N}|u|^{2\alpha\cdot 2^*}dx\right)^{\frac{1}{\tau_2m_2}}\nonumber\\
&=C+\left(\int_{\mathbb{R}^N}|V_1|^qdx\right)^{\frac{1}{q}}\left(\int_{\mathbb{R}^N}|u_0|^2dx\right)^{\frac{\theta}{\tau_1}+\frac{1}{\tau_2m_1}}
\left(\int_{\{|u|\leq 1\}}|u|^{2\alpha\cdot 2^*}dx+\int_{\{|u|>1\}}|u|^{2\alpha\cdot 2^*}dx\right)^{\frac{1}{\tau_2m_2}}\nonumber\\
&\leq C+\left(\int_{\mathbb{R}^N}|V_1|^qdx\right)^{\frac{1}{q}}\left(\int_{\mathbb{R}^N}|u_0|^2dx\right)^{\frac{\theta}{\tau_1}+\frac{1}{\tau_2m_1}}
\left(\int_{\{|u|\leq 1\}}|u|^{2\alpha\cdot 2^*}dx\right)^{\frac{1}{\tau_2m_2}}\nonumber\\
&\quad +\left(\int_{\mathbb{R}^N}|V_1|^qdx\right)^{\frac{1}{q}}\left(\int_{\mathbb{R}^N}|u_0|^2dx\right)^{\frac{\theta}{\tau_1}+\frac{1}{\tau_2m_1}}
\left(\int_{\{|u|>1\}}|u|^{2\alpha\cdot 2^*}dx\right)^{\frac{1}{\tau_2m_2}}\nonumber\\
&\leq C+\left(\int_{\mathbb{R}^N}|V_1|^qdx\right)^{\frac{1}{q}}\left(\int_{\mathbb{R}^N}|u_0|^2dx\right)^{\frac{\theta}{\tau_1}+\frac{1}{\tau_2}}\nonumber\\
&\quad +\left(\int_{\mathbb{R}^N}|V_1|^qdx\right)^{\frac{1}{q}}\left(\int_{\mathbb{R}^N}|u_0|^2dx\right)^{\frac{\theta}{\tau_1}+\frac{1}{\tau_2m_1}}
\left(\int_{\{|u|>1\}}a^{2^*}[|u|+h(|u|^2)]^{2^*}dx\right)^{\frac{1}{\tau_2m_2}}.
\end{align*}
By the mass and energy conversation laws of Lemma 2.1(ii), using Young inequality, we obtain
\begin{align}
&\quad\int_{\mathbb{R}^N}|\nabla u|^2dx+\int_{\mathbb{R}^N}|\nabla h(|u|^2)|^2dx=2E(u_0)+\int_{\mathbb{R}^N}V(x)|u|^2dx\nonumber\\
&\leq C'+\left(\int_{\mathbb{R}^N}|V_1|^qdx\right)^{\frac{1}{q}}\left(\int_{\mathbb{R}^N}|u_0|^2dx\right)^{\frac{\theta}{\tau_1}+\frac{1}{\tau_2m_1}}\nonumber\\
&\quad +\left(\int_{\mathbb{R}^N}|V_1|^qdx\right)^{\frac{1}{q}}\left(\int_{\mathbb{R}^N}|u_0|^2dx\right)^{\frac{\theta}{\tau_1}+\frac{1}{\tau_2m_1}}
\left(\int_{\mathbb{R}^N}a^{2^*}[|u|+h(|u|^2)]^{2^*}dx\right)^{\frac{1}{\tau_2m_2}}\nonumber\\
&\leq C'+\left(\int_{\mathbb{R}^N}|V_1|^qdx\right)^{\frac{1}{q}}\left(\int_{\mathbb{R}^N}|u_0|^2dx\right)^{\frac{\theta}{\tau_1}+\frac{1}{\tau_2m_1}}
 +[a^{2^*}2^{2^*-1}C_s]^{\frac{1}{\tau_2m_2}}\left(\int_{\mathbb{R}^N}|V_1|^qdx\right)^{\frac{1}{q}}\nonumber\\
 &\qquad\qquad\qquad \times\left(\int_{\mathbb{R}^N}|u_0|^2dx\right)^{\frac{\theta}{\tau_1}+\frac{1}{\tau_2m_1}}
\left(\int_{\mathbb{R}^N}[|\nabla u|^2+|\nabla h(|u|^2)|^2]dx\right)^{\frac{2^*}{q(2\alpha\cdot2^*-2)}}\nonumber\\
&\leq C+C'(V,u_0,q)+\frac{1}{2}\int_{\mathbb{R}^N}[|\nabla u|^2+|\nabla h(|u|^2)|^2]dx,\label{1130w2}
\end{align}
which implies that
$$
\int_{\mathbb{R}^N}|\nabla u|^2dx+\int_{\mathbb{R}^N}|\nabla h(|u|^2)|^2dx\leq C.
$$

{\bf Subcase (iv).} $\alpha\geq \frac{2^*+2}{22^*}$, $q>\frac{\alpha\cdot2^*}{(\alpha\cdot2^*-1)}$. Then $\frac{2q}{q-1}<2\alpha\cdot2^*$.

Similar to (\ref{1130w3}), we have
\begin{align}
&\int_{\mathbb{R}^N}|\nabla u|^2dx+\int_{\mathbb{R}^N}|\nabla h(|u|^2)|^2dx\nonumber\\
&\leq C+(a^{2^*}2^{2^*-1} C_s)^{\frac{2}{q(2\alpha\cdot2^*-2)}}C_V\left(\int_{\mathbb{R}^N}|u_0|^2dx\right)^{\frac{[(q-1)(2\alpha\cdot 2^*-2)-2]}{q(2\alpha\cdot2^*-2)}}\nonumber\\
&\qquad \qquad \times\left(\int_{\mathbb{R}^N}|\nabla u|^2dx+\int_{\mathbb{R}^N}|\nabla h(|u|^2)|^2dx\right)^{\frac{2^*}{q(2\alpha\cdot 2^*-2)}}.\label{1130w3'}
\end{align}
Since $\frac{2^*}{q(2\alpha\cdot2^*-2)}<1$ in this subcase, using Young's inequality, we can get
$$
\int_{\mathbb{R}^N}|\nabla u|^2dx+\int_{\mathbb{R}^N}|\nabla h(|u|^2)|^2dx\leq C+\frac{1}{2}\int_{\mathbb{R}^N}|\nabla u|^2dx+\frac{1}{2}\int_{\mathbb{R}^N}|\nabla h(|u|^2)|^2dx,
$$
which means that
$$
\int_{\mathbb{R}^N}|\nabla u|^2dx+\int_{\mathbb{R}^N}|\nabla h(|u|^2)|^2dx\leq C.
$$

{\bf Subcase (v).} $\alpha\geq \frac{2^*+2}{22^*}$, $q=\frac{\alpha\cdot 2^*}{(\alpha\cdot2^*-1)}$. Then $\frac{2q}{q-1}=2\alpha\cdot 2^*$.

Similar to (\ref{3141}), since $\frac{2^*}{q(2\alpha\cdot2^*-2)}<1$ in this subcase, using Young's inequality, we obtain
\begin{align}
&\quad \int_{\mathbb{R}^N}|\nabla u|^2dx+\int_{\mathbb{R}^N}|\nabla h(|u|^2)|^2dx\leq C'+\left(\int_{\mathbb{R}^N}|V_1(x)|^qdx\right)^{\frac{1}{q}}\left(\int_{\mathbb{R}^N}|u|^{\frac{2q}{q-1}}dx\right)^{\frac{q-1}{q}}\nonumber\\
&=C'+\left(\int_{\mathbb{R}^N}|V_1(x)|^{\frac{N}{2}}dx\right)^{\frac{2}{N}}\left(\int_{\mathbb{R}^N}|u^{2\alpha}|^{2^*}dx\right)^{\frac{1}{\alpha\cdot 2^*}}\nonumber\\
&\leq C'+a^2(2^{2^*-1}C_s)^{\frac{1}{\alpha\cdot 2^*}}\left(\int_{\mathbb{R}^N}|V_1(x)|^{\frac{N}{2}}dx\right)^{\frac{2}{N}}
\left(\int_{\mathbb{R}^N}[|\nabla u|^2+|\nabla h(|u|^2)|^2]dx\right)^{\frac{1}{2\alpha}}\nonumber\\
&\leq C+\frac{1}{2}\int_{\mathbb{R}^N}[|\nabla u|^2+|\nabla h(|u|^2)|^2]dx,
\end{align}
 which implies that
$$
\int_{\mathbb{R}^N}|\nabla u|^2dx+\int_{\mathbb{R}^N}|\nabla h(|u|^2)|^2dx\leq C.
$$

Combing the results in subcases above, we prove Theorem 2. \hfill $\Box$

As a corollary of Theorem 1 and Theorem 2, we have the following proposition.

{\bf Proposition 3.1.} {\it Assume that $u$ is the solution of the following problem
\begin{equation}
\label{mhs2} \left\{
\begin{array}{lll}
iu_t=\Delta u+2b^2\alpha |u|^{2\alpha-2}u \Delta (|u|^{2\alpha})+\frac{1}{|x|^m}u \quad {\rm for} \ x\in \mathbb{R}^N\setminus\{\mathbf{0}\}, \ t>0\\
u(x,0)=u_0(x),\quad x\in \mathbb{R}^N.
\end{array}\right.
\end{equation}
Then

1. $b=0$, the solution is global existence for any initial $u_0$ satisfying $0<E(u_0)<+\infty$ if $m<2$ and the solution will blow up in finite time for initial $u_0$ satisfying $-\infty<E(u_0)<0$ if $m\geq 2$.

2. $b>0$ and $0<\alpha<\frac{N-1}{N}$.

(i). $0<\alpha\leq \frac{1}{2}$, the solution is global existence for any initial $u_0$ satisfying $0<E(u_0)<+\infty$ if $m<2$;

(ii). $\frac{1}{2}<\alpha <\frac{N-1}{N}$, the solution is global existence for any initial $u_0$ satisfying $0<E(u_0)<+\infty$ if $$m<\frac{N[2\alpha\cdot 2^*-2]}{2^*},$$
 while
the solution will blow up in finite time for initial $u_0$ satisfying $-\infty<E(u_0)<0$, $xu_0\in L^2(\mathbb{R}^N)$ and $\Im \int_{\mathbb{R}^N}\bar{u_0}(x\cdot \nabla u_0)dx\geq 0$ if $m\geq \frac{N[2\alpha\cdot 2^*-2]}{2^*}$;}

(iii). $\alpha\geq \frac{N-1}{N}$,  the solution is global existence for any initial $u_0$ if $m<\frac{N[\alpha\cdot 2^*-1]}{\alpha\cdot 2^*}$.

{\bf Proof:} We only need to verify the assumptions of Theorem 1 and Theorem 2.

 1. If $b=0$, taking $k=-\frac{1}{2}$, the condition $[\max((2k+1)N,0)+2]V+(x\cdot \nabla V)\leq 0$ in Theorem 1 implies that $2V+(x\cdot \nabla V)\leq 0$, i.e. $m\geq 2$. On the other hand, the condition $q>\frac{2^*}{2^*-2}=\frac{N}{2}$ in Theorem 2 implies that $m<2$.

2. If $b>0$ and $0<\alpha<\frac{N-1}{N}$, taking $k=\alpha-1$ in the condition $$[\max((2k+1)N,0)+2]V+(x\cdot \nabla V)\leq 0$$ in Theorem 1 implies that $m\geq \frac{N[\max(2\alpha,1)\cdot 2^*-2]}{2^*}$. On the other hand, the condition $q>\frac{2^*}{[\max(2\alpha,1)\cdot 2^*-2]}$ in Theorem 2 implies that $m< \frac{N[\max(2\alpha,1)\cdot 2^*-2]}{2^*}$.

3. If $b>0$ and $\alpha\geq \frac{N-1}{N}$, the condition $q>\frac{\alpha\cdot 2^*}{[\alpha\cdot 2^*-1]}$ in Theorem 2 implies that $m<\frac{N[\alpha\cdot 2^*-1]}{\alpha\cdot 2^*}$, then the solution is global existence for any initial $u_0$.\hfill $\Box$

{\bf Remark 3.1.} It is strange that, if $h(s)=0$ or $h(s)=s^{\alpha}$, $0<\alpha\leq \frac{1}{2}$, $q=\frac{N}{2}$, whether the solution is global existence or not just depends on $\|V_1\|_{L^q}$, but independent of $\|u_0\|_{L^2}^2$.

{\bf Remark 3.2.} It is surprised that, in some cases, the watershed of $V(x)$ in this paper meets with that of $W(x)$ in Hartree type nonlinear term in \cite{SW2}.

\section{The Pseudo-conformal Conservation Laws and Asymptotic Behavior for the Solution}
\qquad After obtaining the conditions on the global existence and blowup in finite time, we would like to consider other properties for the solution. Inspired by \cite{Ginibre1, Ginibre2}, we establish pseudo-conformal conservation laws as follows.

{\bf Theorem 3.( Pseudo-conformal Conservation Laws)} {\it 1. Assume that $u$ is the global solution of (\ref{1}), $u_0\in X$ and $xu_0\in L^2(\mathbb{R}^N)$.  Then
\begin{align}
P(t)&=\int_{\mathbb{R}^N}|(x-2it\nabla)u|^2dx+4t^2\int_{\mathbb{R}^N}|\nabla h(|u|^2)|^2dx-4t^2\int_{\mathbb{R}^N}V(x)|u|^2dx\nonumber\\
&=\int_{\mathbb{R}^N}|xu_0|^2dx+4\int_0^t\tau\theta(\tau)d\tau.\label{691}
\end{align}

2.  Assume that $u$ is the blowup solution of (\ref{1}) with blowup time $T$, $u_0\in X$ and $xu_0\in L^2(\mathbb{R}^N)$. Then

\begin{align}
B(t)&:=\int_{\mathbb{R}^N}|(x+2i(T-t)\nabla)u|^2dx+4(T-t)^2\int_{\mathbb{R}^N}|\nabla h(|u|^2)|^2dx\nonumber\\
&\quad-4(T-t)^2\int_{\mathbb{R}^N}V(x)|u|^2dx\nonumber\\
&=\int_{\mathbb{R}^N}|(x+2iT\nabla)u_0|^2dx+4T^2\int_{\mathbb{R}^N}|\nabla h(|u_0|^2)|^2dx-4T^2\int_{\mathbb{R}^N}V(x)|u_0|^2dx\nonumber\\
&\quad-4\int_0^t(T-\tau)\theta(\tau)d\tau.\label{893}
\end{align}

Here
\begin{align}
\theta(t)&=\int_{\mathbb{R}^N}-4N[2h''(|u|^2)h'(|u|^2)|u|^2+( h'(|u|^2))^2]|u|^2|\nabla u|^2dx\nonumber\\
&\qquad-\int_{\mathbb{R}^N} [2V+(x\cdot \nabla V)]|u|^2dx.\label{691'}
\end{align}

}

{\bf Proof of Theorem 3:} 1. Assume that $u$ is the global solution of (\ref{1}), $u_0\in X$ and $xu_0\in L^2(\mathbb{R}^N)$. Using energy conservation law,
we have
\begin{align}
P(t)&=\int_{\mathbb{R}^N}|xu|^2dx+4t\Im \int_{\mathbb{R}^N}\bar{u}(x\cdot \nabla u)dx+4t^2\int_{\mathbb{R}^N}|\nabla u|^2dx\nonumber\\
&\qquad+4t^2\int_{\mathbb{R}^N}|\nabla h(|u|^2)|^2dx-4t^2\int_{\mathbb{R}^N}V(x)|u|^2dx\nonumber\\
&=\int_{\mathbb{R}^N}|xu|^2dx+4t\Im \int_{\mathbb{R}^N}\bar{u}(x\cdot \nabla u)dx+8t^2E(u_0).\label{692}
\end{align}
Noticing that
$$\frac{d}{dt} \int_{\mathbb{R}^N}|x|^2|u|^2dx=-4\Im \int_{\mathbb{R}^N} \bar{u}(x\cdot \nabla u)dx,$$
we obtain
\begin{align}
P'(t)&=\frac{d}{dt}\int_{\mathbb{R}^N}|xu|^2dx+4\Im \int_{\mathbb{R}^N}\bar{u}(x\cdot \nabla u)dx+4t\frac{d}{dt}\Im \int_{\mathbb{R}^N}\bar{u}(x\cdot \nabla u)dx+16tE(u_0)\nonumber\\
&=4t\frac{d}{dt}\Im \int_{\mathbb{R}^N}\bar{u}(x\cdot \nabla u)dx+16tE(u_0)\nonumber\\
&=4t\left\{-2\int_{\mathbb{R}^N}|\nabla u|^2dx-(N+2)\int_{\mathbb{R}^N}|\nabla h(|u|^2)|^2dx\right.\nonumber\\
&\qquad \left.-8N\int_{\mathbb{R}^N}h''(|u|^2)h'(|u|^2)|u|^4|\nabla u|^2dx-\int_{\mathbb{R}^N}(x\cdot \nabla V)|u|^2dx\right\}\nonumber\\
&\qquad+8t\int_{\mathbb{R}^N}[|\nabla u|^2+|\nabla h(|u|^2)|^2]dx-8t\int_{\mathbb{R}^N}V(x)|u|^2dx\nonumber\\
&=4t\int_{\mathbb{R}^N}-4N[2h''(|u|^2)h'(|u|^2)|u|^2+(h'(|u|^2))^2]|u|^2|\nabla u|^2dx\nonumber\\
&\quad-4t\int_{\mathbb{R}^N}[2V(x)+(x\cdot \nabla V)]|u|^2dx\nonumber\\
&=4t\theta(t).\label{693}
\end{align}
Integrating (\ref{693}) from $0$ to $t$, we have
\begin{align}
P(t)&=\int_{\mathbb{R}^N}|(x-2it\nabla)u|^2dx+4t^2\int_{\mathbb{R}^N}|\nabla h(|u|^2)|^2dx-4t^2\int_{\mathbb{R}^N}V(x)|u|^2dx\nonumber\\
&=\int_{\mathbb{R}^N}|xu_0|^2dx+4\int_0^t\tau\theta(\tau)d\tau,\label{694}
\end{align}
where $\theta(\tau)$ is defined by (\ref{691'}).

2. Assume that $u$ is the blowup solution of (\ref{1}),  $u_0\in X$ and $xu_0\in L^2(\mathbb{R}^N)$. By energy conservation law, we get
\begin{align}
B(t)&:=\int_{\mathbb{R}^N}|(x+2i(T-t)\nabla)u|^2dx+4(T-t)^2\int_{\mathbb{R}^N}|\nabla h(|u|^2)|^2dx\nonumber\\
&\quad-4(T-t)^2\int_{\mathbb{R}^N}V(x)|u|^2dx\nonumber\\
&=\int_{\mathbb{R}^N}|xu|^2dx-4(T-t)\Im \int_{\mathbb{R}^N}\bar{u}(x\cdot \nabla u)dx+4(T-t)^2\int_{\mathbb{R}^N}|\nabla u|^2dx\nonumber\\
&\qquad+4(T-t)^2\int_{\mathbb{R}^N}|\nabla h(|u|^2)|^2dx-4(T-t)^2\int_{\mathbb{R}^N}V(x)|u|^2dx\nonumber\\
&=\int_{\mathbb{R}^N}|xu|^2dx-4(T-t)\Im \int_{\mathbb{R}^N}\bar{u}(x\cdot \nabla u)dx+8(T-t)^2E(u_0)\label{891}
\end{align}
and
\begin{align}
B'(t)&=\frac{d}{dt}\int_{\mathbb{R}^N}|xu|^2dx+4\Im \int_{\mathbb{R}^N}\bar{u}(x\cdot \nabla u)dx\nonumber\\
&\quad-4(T-t)\frac{d}{dt}\Im \int_{\mathbb{R}^N}\bar{u}(x\cdot \nabla u)dx-16(T-t)E(u_0)\nonumber\\
&=-4(T-t)\frac{d}{dt}\Im \int_{\mathbb{R}^N}\bar{u}(x\cdot \nabla u)dx-16(T-t)E(u_0)\nonumber\\
&=4(T-t)\left\{\int_{\mathbb{R}^N}4N[2h''(|u|^2)h'(|u|^2)|u|^2+( h'(|u|^2))^2]|u|^2|\nabla u|^2dx\right.\nonumber\\
&\qquad\left.+\int_{\mathbb{R}^N} [2V(x)+(x\cdot \nabla V)]|u|^2dx\right\}.\label{892}
\end{align}
Integrating (\ref{892}) from $0$ to $t$, we have
\begin{align*}
B(t)&=B(0)+32E(u_0)\int_0^t(T-\tau)d\tau-4\int_0^t(T-\tau)\theta(\tau)d\tau\nonumber\\
&=\int_{\mathbb{R}^N}|(x+2iT\nabla)u_0|^2dx+4T^2\int_{\mathbb{R}^N}|\nabla h(|u_0|^2)|^2dx\nonumber\\
&\quad-4T^2\int_{\mathbb{R}^N}V(x)|u_0|^2dx-4\int_0^t(T-\tau)\theta(\tau)d\tau,
\end{align*}
where $\theta(\tau)$ is defined by (\ref{691'}).\hfill $\Box$

As the application of Theorem 3, we have a theorem as follows.

{\bf Theorem 4. (Asymptotic behaviors for the solution)} {\it 1. Assume that $u$ is the global solution of (\ref{1}), $u_0\in X$, $xu_0\in L^2(\mathbb{R}^N)$, $V(x)=V_1(x)+V_2(x)\in L^q(\mathbb{R}^N)+L^{\infty}(\mathbb{R}^N)$ for some $q>1$ and $V(x)\leq 0$ for $x\in \mathbb{R}^N$. Then the following properties hold:

(1) If $2h''(s)h'(s)s+(h'(s))^2\geq 0$ for $s\geq 0$, and $2V+(x\cdot \nabla V)\geq 0$ for $x\in \mathbb{R}^N$, then
\begin{align}
\int_{\mathbb{R}^N}|\nabla h(|u|^2)|^2dx +\int_{\mathbb{R}^N}|V(x)||u|^2dx\leq \frac{C}{t^2}\quad {\rm for}\ t\geq 1. \label{6101}
\end{align}

(2) If $2h''(s)h'(s)s+(h'(s))^2\geq 0$ for $s\geq 0$, and there exists $0<c<2$ such that $-c|V|\leq 2V+(x\cdot \nabla V)\leq 0$ for $x\in \mathbb{R}^N$, then
\begin{align}
\int_{\mathbb{R}^N}|\nabla h(|u|^2)|^2dx +\int_{\mathbb{R}^N}|V(x)||u|^2dx\leq\frac{C}{t^{2-c}}\quad {\rm for}\ t\geq 1.\label{6102}
\end{align}

(3) If there exists $0<k_1<\frac{2}{N}$ such that $-k_1(h'(s))^2\leq 2h''(s)h'(s)s+(h'(s))^2\leq 0$ for $s\geq 0$, and $2V+(x\cdot \nabla V)\geq 0$ for $x\in \mathbb{R}^N$, then
\begin{align}
\int_{\mathbb{R}^N}|\nabla h(|u|^2)|^2dx +\int_{\mathbb{R}^N}|V(x)||u|^2dx\leq \frac{C}{t^{2-Nk_1}}\quad {\rm for}\ t\geq 1.\label{6103}
\end{align}

(4) If there exist $0<k_1<\frac{2}{N}$ and $0<c<2$ such that $-k_1(h'(s))^2\leq 2h''(s)h'(s)s+(h'(s))^2\leq 0$ for $s\geq 0$, and $-c|V|\leq 2V+(x\cdot \nabla V)\leq 0$ for $x\in \mathbb{R}^N$, then
\begin{align}
\int_{\mathbb{R}^N}|\nabla h(|u|^2)|^2dx +\int_{\mathbb{R}^N}|V(x)||u|^2dx\leq \frac{C}{t^{2-\max(Nk_1,c)}}\quad {\rm for}\ t\geq 1.\label{6103}
\end{align}

In all cases above, by the conservation of energy, we have
\begin{align}
\lim_{t\rightarrow +\infty} \int_{\mathbb{R}^N}|\nabla u(x,t)|^2dx=2E(u_0),\quad \lim_{t\rightarrow +\infty} \|u(\cdot,t)\|_{H^1}^2=M(u_0)+2E(u_0).\label{9181}
\end{align}

2. Assume that $u$ is the blowup solution of (\ref{1}), $[(h'(s))^2+2h''(s)h'(s)s]\leq 0$ for $s\geq 0$ and $2V+(x\cdot \nabla V)\leq 0$ for $x\in \mathbb{R}^N$, $u_0\in X$ and $xu_0\in L^2(\mathbb{R}^N)$. If $E(u_0)\leq 0$ and $-4T^2E(u_0)-\int_{\mathbb{R}^N}|xu_0|^2dx-4T\Im\int_{\mathbb{R}^N}\bar{u}_0(x\cdot \nabla u_0)dx>0$, then
\begin{align}
\int_{\mathbb{R}^N}[|\nabla u|^2+|\nabla h(|u|^2)|^2]dx\geq \frac{C}{(T-t)^2},\quad \int_{\mathbb{R}^N}V(x)|u|^2dx\geq \frac{C}{(T-t)^2}.\label{895}
\end{align}
}

{\bf Proof of Theorem 4:} 1. Assume that $u$ is the global solution of (\ref{1}), $u_0\in X$ and $xu_0\in L^2(\mathbb{R}^N)$, $V(x)\leq 0$.

(1) $2h''(s)h'(s)s+(h'(s))^2\geq 0$ and $2V+(x\cdot \nabla V)\geq 0$. (\ref{691}) implies that
$$
4t^2\int_{\mathbb{R}^N}|\nabla h(|u|^2)|^2dx+4t^2\int_{\mathbb{R}^N}|V(x)||u|^2dx\leq\int_{\mathbb{R}^N}|xu_0|^2dx,
$$
i.e.,
$$
\int_{\mathbb{R}^N}|\nabla h(|u|^2)|^2dx+\int_{\mathbb{R}^N}|V(x)||u|^2dx\leq \frac{C}{t^2}.
$$

(2) $2h''(s)h'(s)s+(h'(s))^2\geq 0$ and $-c|V|\leq 2V+(x\cdot \nabla V)\leq 0$ for some $0<c<2$. (\ref{691}) means that
\begin{align}
&\quad 4t^2\int_{\mathbb{R}^N}|\nabla h(|u|^2)|^2dx+4t^2\int_{\mathbb{R}^N}|V(x)||u|^2dx\nonumber\\
&\leq\int_{\mathbb{R}^N}|xu_0|^2dx+4c\int_0^t\tau\left(\int_{\mathbb{R}^N}|V(x)||u|^2dx\right)d\tau.
\label{6232}
\end{align}
Let
$$
A_1(t):=4\int_0^t\tau\left(\int_{\mathbb{R}^N}|V(x)||u|^2dx\right)d\tau.
$$
(\ref{6232}) implies
$$
A'_1(t)\leq \frac{C_0}{t}+\frac{c}{t}A_1(t).
$$
Using Gronwll's inequality, we have
$$
A_1(t)\leq t^c[A_1(1)+C-\frac{C}{t^c}]\leq C't^c.
$$
Consequently,
$$
\int_{\mathbb{R}^N}|\nabla h(|u|^2)|^2dx+\int_{\mathbb{R}^N}|V(x)||u|^2dx\leq \frac{C}{t^{2-c}}.
$$

(3) $-k_1(h'(s))^2\leq 2h''(s)h'(s)s+(h'(s))^2<0$ for some $0<k_1<\frac{2}{N}$ and $2V+(x\cdot \nabla V)\geq 0$.
(\ref{691}) means that
\begin{align}
&\quad 4t^2\int_{\mathbb{R}^N}|\nabla h(|u|^2)|^2dx+4t^2\int_{\mathbb{R}^N}|V(x)||u|^2dx\nonumber\\
&\leq\int_{\mathbb{R}^N}|xu_0|^2dx+4Nk_1\int_0^t\tau\left(\int_{\mathbb{R}^N}|\nabla h(|u|^2)|^2dx\right)d\tau.
\label{914w1}
\end{align}
Let
$$
A_2(t):=4\int_0^t\tau\left(\int_{\mathbb{R}^N}|\nabla h(|u|^2)|^2dx\right)d\tau.
$$
(\ref{914w1}) equals to
$$
A'_2(t)\leq \frac{C_0}{t}+\frac{Nk_1}{t}A_2(t).
$$
Using Gronwell's inequality, we get
$$
A_2(t)\leq t^{Nk_1}c[A_2(1)+C-\frac{C}{t^{Nk_1}}]\leq C't^{Nk_1}.
$$
Consequently, we have
$$
\int_{\mathbb{R}^N}|\nabla h(|u|^2)|^2dx+\int_{\mathbb{R}^N}|V(x)||u|^2dx\leq \frac{C}{t^{2-Nk_1}}.
$$

(4) $-k_1(h'(s))^2\leq 2h''(s)h'(s)s+(h'(s))^2<0$ for some $0<k_1<\frac{2}{N}$ and $-c|V|\leq 2V+(x\cdot \nabla V)\leq 0$ for some $0<c<2$.
(\ref{691}) means that
\begin{align}
&\quad 4t^2\int_{\mathbb{R}^N}|\nabla h(|u|^2)|^2dx+4t^2\int_{\mathbb{R}^N}|V(x)||u|^2dx\nonumber\\
&\leq\int_{\mathbb{R}^N}|xu_0|^2dx+4Nk_1\int_0^t\tau\left(\int_{\mathbb{R}^N}|\nabla h(|u|^2)|^2dx\right)d\tau
+4c\int_0^t\tau\left(\int_{\mathbb{R}^N}|V(x)||u|^2dx\right)d\tau\nonumber\\
&\leq C+4\max(Nk_1,c)\int_0^t\tau\left[\int_{\mathbb{R}^N}|\nabla h(|u|^2)|^2dx+\int_{\mathbb{R}^N}|V(x)||u|^2dx\right]d\tau.\label{914w2}
\end{align}
Let
$$
A_3(t):=4\int_0^t\tau\left[\int_{\mathbb{R}^N}|\nabla h(|u|^2)|^2dx+\int_{\mathbb{R}^N}|V(x)||u|^2dx\right]d\tau.
$$
(\ref{914w2}) implies
$$
A'_3(t)\leq \frac{C_0}{t}+\frac{\max(Nk_1,c)}{t}A_3(t).
$$
Using Gronwell's inequality, we get
$$
A_3(t)\leq t^{\max(Nk_1,c)}[A_3(1)+C-\frac{C}{t^{\max(Nk_1,c)}}]\leq C't^{\max(Nk_1,c)}.
$$
Consequently, we obtain
$$
\int_{\mathbb{R}^N}|\nabla h(|u|^2)|^2dx+\int_{\mathbb{R}^N}|V(x)||u|^2dx\leq \frac{C}{t^{2-\max(Nk_1,c)}}.
$$

In all cases above, we have
$$
\lim_{t\rightarrow +\infty}\int_{\mathbb{R}^N}|\nabla h(|u|^2)|^2dx+\int_{\mathbb{R}^N}|V(x)||u|^2dx=0.
$$
By the conservation of energy, we get
$$
\lim_{t\rightarrow +\infty}\left(\frac{1}{2}\int_{\mathbb{R}^N}|\nabla u|^2dx+\frac{1}{2}\int_{\mathbb{R}^N}|\nabla h(|u|^2)|^2dx+\frac{1}{2}\int_{\mathbb{R}^N}|V(x)||u|^2dx\right)=E(u_0),
$$
which means that
$$
\lim_{t\rightarrow +\infty}\int_{\mathbb{R}^N}|\nabla u|^2dx=2E(u_0).
$$
By the conservation of mass, we obtain
$$
\lim_{t\rightarrow +\infty}\|u(\cdot,t)\|_{H^1}^2=\lim_{t\rightarrow +\infty}\left(\int_{\mathbb{R}^N}|u|^2dx+\int_{\mathbb{R}^N}|\nabla u|^2dx\right)=M(u_0)+2E(u_0).
$$
(\ref{9181}) is proved.

2. Assume that $u$ is the blowup solution of (\ref{1}),  $u_0\in X$ and $xu_0\in L^2(\mathbb{R}^N)$, $V(x)\geq 0$ and $2V+(x\cdot \nabla V)\leq 0$, $2h''(s)h'(s)s+(h'(s))^2\leq 0$. Using (\ref{893}), we have
\begin{align}
&\quad 2(T-t)^2\int_{\mathbb{R}^N}V(x)|u|^2dx\nonumber\\
&=\int_{\mathbb{R}^N}|(x+2i(T-t)\nabla)u|^2dx+4(T-t)^2\int_{\mathbb{R}^N}|\nabla h(|u|^2)|^2dx\nonumber\\
&\quad+4\int_0^t(T-\tau)\theta(\tau)d\tau-8T^2E(u_0)-\int_{\mathbb{R}^N}|xu_0|^2dx\nonumber\\
&\quad-4T\Im\int_{\mathbb{R}^N}\bar{u}_0(x\cdot \nabla u_0)dx.\label{894}
\end{align}
If $$-8T^2E(u_0)-\int_{\mathbb{R}^N}|xu_0|^2dx-4T\Im\int_{\mathbb{R}^N}\bar{u}_0(x\cdot \nabla u_0)dx>0,$$ then (\ref{894}) implies that
\begin{align*}
\int_{\mathbb{R}^N}V(x)|u|^2dx\geq \frac{C}{(T-t)^2}.
\end{align*}

Using energy conservation law $E(u)=E(u_0)$, we get
$$
\frac{1}{2}\int_{\mathbb{R}^N}[|\nabla u|^2+|\nabla h(|u|^2)|^2]dx=\frac{1}{2}\int_{\mathbb{R}^N}V(x)|u|^2dx+E(u_0)\geq
\frac{C}{(T-t)^2}+E(u_0).
$$
As $t$ close to $T$ enough, we have
$$
\frac{C}{(T-t)^2}+E(u_0)\geq \frac{C'}{(T-t)^2}.
$$
for some constant $0<C'<C$. Hence
$$
\int_{\mathbb{R}^N}|\nabla u|^2dx+\int_{\mathbb{R}^N}|\nabla h(|u|^2)|^2dx\geq \frac{2C'}{(T-t)^2},
$$
(\ref{895}) holds.\hfill $\Box$

We would like to give two examples to illustrate the results intuitively.

{\bf Example 4.1.}
\begin{equation}
\label{mhs312} \left\{
\begin{array}{lll}
iu_t=\Delta u+2b^2\alpha |u|^{2\alpha-2}u \Delta (|u|^{2\alpha})-\frac{1}{|x|^m}u, \quad  x\in \mathbb{R}^N\setminus\{\mathbf{0}\}, \ t>0\\
u(x,0)=u_0(x),\quad x\in \mathbb{R}^N.
\end{array}\right.
\end{equation}
Since $h(s)=s^{\alpha}$ and $V(x)=-\frac{1}{|x|^m}(x\neq \mathbf{0})$, $2h''(s)h'(s)s+(h'(s))^2\geq 0$ and $2V+(x\cdot \nabla V)\geq 0$ imply that
$\alpha\geq \frac{1}{2}$ and $m\geq 2$, while $-k_1(h'(s))^2\leq 2h''(s)h'(s)s+(h'(s))^2<0$ and $-c|V|\leq 2V+(x\cdot \nabla V)<0$
imply that $\alpha<\frac{1}{2}$ and $m<2$. We can take $k_1=1-2\alpha$, $c=2-m$ and get
$$
\int_{\mathbb{R}^N}|\nabla |u|^{2\alpha}|^2dx +\int_{\mathbb{R}^N}\frac{1}{|x|^m}|u|^2dx\leq \frac{C}{t^l}\quad {\rm for}\ t\geq 1.
$$
Here $l=2$ or $l=m$ if $\alpha\geq \frac{1}{2}$, $l=2-N(1-2\alpha)$ or $l=\max(m, 2-N(1-2\alpha))$ if $\alpha<\frac{1}{2}$.

{\bf Example 4.2.} Noticing that $h(s)=s^{\alpha}$ and $V(x)=\frac{1}{|x|^m}(x\neq \mathbf{0})$ in (\ref{mhs2}),
 $2h''(s)h'(s)s+(h'(s))^2\leq 0$ and $2V+(x\cdot \nabla V)\leq 0$ imply that $\alpha\leq \frac{1}{2}$ and $m\geq 2$.
 And
 $$
 \int_{\mathbb{R}^N}[|\nabla u|^2+|\nabla |u|^{2\alpha}|^2]dx\geq \frac{C}{(T-t)^2},\quad \int_{\mathbb{R}^N}\frac{1}{|x|^m}|u|^2dx\geq \frac{C}{(T-t)^2}.
 $$

Basing on Theorem 3 and Theorem 4, we can obtain Morawetz estimates and spacetime bounds for the solution of (\ref{1}) below.

{\bf Theorem 5.((Morawetz estimates and Spacetime bounds for the global solution)} {\it Assume that $u(x,t)$ is the global solution of (\ref{1}), $u_0\in X$, $xu_0\in L^2(\mathbb{R}^N)$, $0<E(u_0)<+\infty$, $V(x)=V_1(x)+V_2(x)\in L^q(\mathbb{R}^N)+L^{\infty}(\mathbb{R}^N)$ for some $q>1$ and $V(x)\leq 0$ for $x\in \mathbb{R}^N$.
Then

1. Morawetz estimates

If $a(x,t)\geq at^{\lambda}$ for some $a>0$ and $\lambda<1$ when $0\leq t\leq 1$,
$a(x,t)\geq bt^{\mu}$ for some $b>0$ and $\mu+l>1$ when $t\geq 1$, then
\begin{align}
\int_0^{+\infty}\int_{\mathbb{R}^N}\frac{[|\nabla h(|u|^2)|^2+|V(x)||u|^2]}{a(x,t)}dxdt\leq C;\label{491}
\end{align}
Especially, if $a(x,t)=(|x|+t)^{\lambda}$, $0\leq \lambda<1$, $\lambda+l>1$, then
\begin{align}
\int_0^{+\infty}\int_{\mathbb{R}^N}\frac{[|\nabla h(|u|^2)|^2+|V(x)||u|^2]}{(|x|+t)^{\lambda}}dxdt\leq C.\label{4201}
\end{align}
If $a(x,t)\equiv a>0$, $l>1$, then
\begin{align}
\int_0^{+\infty}\int_{\mathbb{R}^N}[|\nabla h(|u|^2)|^2+|V(x)||u|^2]dxdt\leq C.\label{4202}
\end{align}

2. Spacetime bounds

(1). For $Ml>1$, we have
\begin{align}
\left(\int_0^{+\infty}\left(\int_{\mathbb{R}^N}[|\nabla h(|u|^2)|^2+|V(x)||u|^2]dx\right)^Mdt\right)^{\frac{1}{M}}\leq C
\end{align}
in one of the following cases: (i) Case (1) of Theorem 4;(ii) $0<c<1$ in Case (2) of Theorem 4; (iii) $0<k_1<\frac{1}{N}$ in Case (3) of Theorem 4;
(iv) $0<c<1$, $0<k_1<\frac{1}{N}$ in Case (4) of Theorem 4.

(2). If there exist $m_1>1$, $m_2>1$, $c_1>0$ and $c_2>0$ such that $|u|^{m_1\bar{r}}\leq c_1[h(|u|^2)]^{2^*}$ for $0\leq s\leq 1$ and $|u|^{m_2\bar{r}}\leq c_2[h(|u|^2)]^{2^*}$ for $s>1$, then
\begin{align}
\|u\|_{L^{\bar{q}}_tL^{\bar{r}}_x}=\left(\int_0^{+\infty}\left(\int_{\mathbb{R}^N}|u|^{\bar{r}}dx\right)^{\frac{\bar{q}}{\bar{r}}}dt\right)^{\frac{1}{\bar{q}}}\leq C\label{4103}
\end{align}
in one of the cases of (i)--(iv). Here
\begin{align}
\bar{r}>2,\quad q>\frac{2\bar{r}(m_1\bar{r}-2)}{2^*l(\bar{r}-2)},\qquad q>\frac{2\bar{r}(m_2\bar{r}-2)}{2^*l(\bar{r}-2)}.\label{410w1}
\end{align}
 }

{\bf Proof:} 1. By energy conservation law,
\begin{align}
&\quad\int_{\mathbb{R}^N}[|\nabla h(|u|^2)|^2+|V(x)||u|^2]dx\leq \int_{\mathbb{R}^N}[|\nabla u|^2+|\nabla h(|u|^2)|^2+|V(x)||u|^2]dx\nonumber\\
&\leq 2E(u_0)\quad {\rm for\ all}\ t\geq 0,\quad {\rm especially\ for}\ 0\leq t\leq 1.\label{4101}
\end{align}
On the other hand, by the proof of Theorem 4,
\begin{align}
\int_{\mathbb{R}^N}[|\nabla h(|u|^2)|^2+|V(x)||u|^2]dx&\leq \frac{C}{t^l}\quad {\rm for}\quad t\geq 1.\label{4102}
\end{align}
Here $l=2$ in Case (1) of Theorem 4; $l=2-c$ in Case (2) of Theorem 4; $l=2-Nk_1$ in Case (3) of Theorem 4; $l=2-\max(Nk_1,c)$ in Case (4) of Theorem 4.
Consequently,
\begin{align}
&\quad\int_0^{+\infty}\int_{\mathbb{R}^N}\frac{[|\nabla h(|u|^2)|^2+|V(x)||u|^2]}{a(x,t)}dxdt\nonumber\\
&=\int_0^1\int_{\mathbb{R}^N}\frac{[|\nabla h(|u|^2)|^2+|V(x)||u|^2]}{a(x,t)}dxdt+
\int_1^{+\infty}\int_{\mathbb{R}^N}\frac{[|\nabla h(|u|^2)|^2+|V(x)||u|^2]}{a(x,t)}dxdt\nonumber\\
&\leq \int_0^1 \frac{2E(u_0)}{at^{\lambda}}dt+\int_1^{+\infty}\frac{C}{bt^{\mu+l}}dt\leq \frac{2E(u_0)}{a(1-\lambda)}+\frac{C}{b(\mu+l-1)}.
\end{align}
Especially, if $a(x,t)=(|x|+t)^{\lambda}$, $0<\lambda<1$, $\lambda+l>1$, then
\begin{align*}
\int_0^{+\infty}\int_{\mathbb{R}^N}\frac{[|\nabla h(|u|^2)|^2+|V(x)||u|^2]}{(|x|+t)^{\lambda}}dxdt\leq\frac{2E(u_0)}{a(1-\lambda)}+\frac{C}{b(\lambda+l-1)}.
\end{align*}
If $a(x,t)\equiv 1$, $l>1$, then
\begin{align*}
\int_0^{+\infty}\int_{\mathbb{R}^N}[|\nabla h(|u|^2)|^2+|V(x)||u|^2]dxdt\leq C.
\end{align*}

2. (1). If $lM>1$, denote $\phi(u)=\int_{\mathbb{R}^N}[|\nabla h(|u|^2)|^2+|V(x)||u|^2]dx$, we have
 \begin{align}
&\quad\left(\int_0^{+\infty}\left(\int_{\mathbb{R}^N}[|\nabla h(|u|^2)|^2+|V(x)||u|^2]dx\right)^Mdt\right)^{\frac{1}{M}}\nonumber\\
&=\left\{\int_0^1\left(\phi(u)\right)^M+\int_1^{+\infty}\left(\phi(u)\right)^M dt\right\}^{\frac{1}{M}}\nonumber\\
&\leq \left(\int_0^1[2E(u_0)]^Mdt+\int_1^{+\infty}\frac{C}{t^{lM}} dt\right)^{\frac{1}{M}}\leq \left([2E(u_0)]^M+\frac{C}{lM-1}\right)^{\frac{1}{M}}\leq C.
\end{align}

(2). Let
$$
\tau_1=\frac{m_1\bar{r}-2}{(m_1-1)\bar{r}},\quad \tau'_1=\frac{m_1\bar{r}-2}{\bar{r}-2},\quad \tau_2=\frac{m_2\bar{r}-2}{(m_2-1)\bar{r}},\quad \tau'_2=\frac{m_2\bar{r}-2}{\bar{r}-2}.
$$
Note that
\begin{align}
&\quad\int_{\mathbb{R}^N}|u|^{\bar{r}}dx=\int_{\{|u|\leq 1\}}|u|^{\bar{r}}dx+\int_{\{|u|>1\}}|u|^{\bar{r}}dx\nonumber\\
&\leq \left(\int_{\{|u|\leq 1\}}|u|^2dx\right)^{\frac{1}{\tau_1}}\left(\int_{\{|u|\leq 1\}}|u|^{m_1\bar{r}}dx\right)^{\frac{1}{\tau'_1}}\nonumber\\
&\qquad+\left(\int_{\{|u|>1\}}|u|^2dx\right)^{\frac{1}{\tau_2}}\left(\int_{\{|u|> 1\}}|u|^{m_1\bar{r}}dx\right)^{\frac{1}{\tau'_2}}\nonumber\\
&\leq \left(\int_{\mathbb{R}^N}|u|^2dx\right)^{\frac{1}{\tau_1}}\left(\int_{\{|u|\leq 1\}}c_1[h(|u|^2)]^{2^*}dx\right)^{\frac{1}{\tau'_1}}\nonumber\\
&\qquad+\left(\int_{\mathbb{R}^N}|u|^2dx\right)^{\frac{1}{\tau_2}}\left(\int_{\{|u|> 1\}}c_2[h(|u|^2)]^{2^*}dx\right)^{\frac{1}{\tau'_2}}\nonumber\\
&\leq \left(\int_{\mathbb{R}^N}|u|^2dx\right)^{\frac{1}{\tau_1}}\left(\int_{\mathbb{R}^N}c_1[h(|u|^2)]^{2^*}dx\right)^{\frac{1}{\tau'_1}}
+\left(\int_{\mathbb{R}^N}|u|^2dx\right)^{\frac{1}{\tau_2}}\left(\int_{\mathbb{R}^N}c_2[h(|u|^2)]^{2^*}dx\right)^{\frac{1}{\tau'_2}}\nonumber\\
&\leq C\left(\int_{\mathbb{R}^N}|\nabla h(|u|^2)|^2dx\right)^{\frac{2^*}{2\tau'_1}}+C\left(\int_{\mathbb{R}^N}|\nabla h(|u|^2)|^2dx\right)^{\frac{2^*}{2\tau'_2}}.\label{410x1}
\end{align}

If
$$
q>\frac{2\bar{r}(m_1\bar{r}-2)}{2^*l(\bar{r}-2)},\qquad q>\frac{2\bar{r}(m_2\bar{r}-2)}{2^*l(\bar{r}-2)},
$$
by (\ref{4101}), (\ref{4102}) and (\ref{410x1}), then
\begin{align*}
&\quad\|u\|_{L^{\bar{q}}_t(\mathbb{R})L^{\bar{r}}_x(\mathbb{R}^N)}=\left(\int_0^{+\infty}\left(\int_{\mathbb{R}^N}|u|^{\bar{r}}dx\right)^{\frac{\bar{q}}{\bar{r}}}dt\right)^{\frac{1}{\bar{q}}}\nonumber\\
&\leq C\left(\int_0^{+\infty}\left[\left(\int_{\mathbb{R}^N}|\nabla h(|u|^2)|^2dx\right)^{\frac{2^*}{2\tau'_1}}+\left(\int_{\mathbb{R}^N}|\nabla h(|u|^2)|^2dx\right)^{\frac{2^*}{2\tau'_2}}\right]^{\frac{\bar{q}}{\bar{r}}}dt\right)^{\frac{1}{\bar{q}}}\nonumber\\
&\leq C\left(\int_0^{+\infty}\left(\int_{\mathbb{R}^N}|\nabla h(|u|^2)|^2dx\right)^{\frac{\bar{q}\cdot2^*}{2\bar{r}\tau'_1}}dt\right)^{\frac{1}{\bar{q}}}
+C\left(\int_0^{+\infty}\left(\int_{\mathbb{R}^N}|\nabla h(|u|^2)|^2dx\right)^{\frac{\bar{q}\cdot2^*}{2\bar{r}\tau'_2}}dt\right)^{\frac{1}{\bar{q}}}\nonumber\\
&\leq C\left(\int_0^1\left(\int_{\mathbb{R}^N}|\nabla h(|u|^2)|^2dx\right)^{\frac{\bar{q}\cdot2^*}{2\bar{r}\tau'_1}}dt\right)^{\frac{1}{\bar{q}}}
+C\left(\int_1^{+\infty}\left(\int_{\mathbb{R}^N}|\nabla h(|u|^2)|^2dx\right)^{\frac{\bar{q}\cdot2^*}{2\bar{r}\tau'_1}}dt\right)^{\frac{1}{\bar{q}}}\nonumber\\
&\quad+ C\left(\int_0^1\left(\int_{\mathbb{R}^N}|\nabla h(|u|^2)|^2dx\right)^{\frac{\bar{q}\cdot2^*}{2\bar{r}\tau'_2}}dt\right)^{\frac{1}{\bar{q}}}
+C\left(\int_1^{+\infty}\left(\int_{\mathbb{R}^N}|\nabla h(|u|^2)|^2dx\right)^{\frac{\bar{q}\cdot2^*}{2\bar{r}\tau'_2}}dt\right)^{\frac{1}{\bar{q}}}\nonumber\\
&\leq C\left(\int_0^1[2E(u_0)]^{\frac{\bar{q}\cdot2^*}{2\bar{r}\tau'_1}}dt\right)^{\frac{1}{\bar{q}}}+C\left(\int_1^{+\infty}\frac {1}{t^{\frac{l\bar{q}\cdot2^*}{2\bar{r}\tau'_1}}}dt\right)^{\frac{1}{\bar{q}}}\nonumber\\
&\quad + C\left(\int_0^1[2E(u_0)]^{\frac{\bar{q}\cdot2^*}{2\bar{r}\tau'_2}}dt\right)^{\frac{1}{\bar{q}}}+C\left(\int_1^{+\infty}\frac {1}{t^{\frac{l\bar{q}\cdot2^*}{2\bar{r}\tau'_2}}}dt\right)^{\frac{1}{\bar{q}}}\nonumber\\
&\leq C'<+\infty.
\end{align*}
Theorem 5 is proved.\hfill $\Box$

We would like to give an example to illustrate the results of Theorem 5.

{\bf Example 4.3.}
\begin{equation}
\label{mhs410} \left\{
\begin{array}{lll}
iu_t=\Delta u+2[\alpha_1 |u|^{2\alpha_1-2}+\alpha_2 |u|^{2\alpha_2-2}]u \Delta (|u|^{2\alpha_1}+|u|^{2\alpha_2})-\frac{1}{|x|^m}u,\ x\in \mathbb{R}^N\setminus\{\mathbf{0}\}, \ t>0\\
u(x,0)=u_0(x),\quad x\in \mathbb{R}^N.
\end{array}\right.
\end{equation}

1. Since $h(s)=s^{\alpha_1}+s^{\alpha_2}$ and $V(x)=-\frac{1}{|x|^m}$, $2h''(s)h'(s)s+(h'(s))^2\geq 0$ and $2V+(x\cdot \nabla V)\geq 0$ imply that
$\alpha_1, \alpha_2\geq \frac{1}{2}$ and $m\geq 2$. While $-k_1(h'(s))^2\leq 2h''(s)h'(s)s+(h'(s))^2<0$ and $-c|V|\leq 2V+(x\cdot \nabla V)<0$
imply that $\alpha_1, \alpha_2<\frac{1}{2}$ and $m<2$, we can take $k_1=\max(1-2\alpha_1, 1-2\alpha_2)$, $c=2-m$ in this case. Under certain assumptions, we get
$$
\int_0^{+\infty}\int_{\mathbb{R}^N}(|\nabla [|u|^{2\alpha_1}+|u|^{2\alpha_2}]|^2+\frac{1}{|x|^m}|u|^2)dxdt\leq C,
$$
$$
\int_0^{+\infty}\int_{\mathbb{R}^N}\frac{(|\nabla [|u|^{2\alpha_1}+|u|^{2\alpha_2}]|^2+\frac{1}{|x|^m}|u|^2)}{(|x|+t)^{\lambda}}dxdt\leq C.
$$

2. If $1<m_1=\frac{\min(\alpha_1,\alpha_2)2^*}{\bar{r}}$ and $1<m_2=\frac{\max(\alpha_1,\alpha_2)2^*}{\bar{r}}$, then
$$
\|u\|_{L^{\bar{q}}_t(\mathbb{R})L^{\bar{r}}_x(\mathbb{R}^N)}
=\left(\int_0^{+\infty}\left(\int_{\mathbb{R}^N}|u|^{\bar{r}}dx\right)^{\frac{\bar{q}}{\bar{r}}}dt\right)^{\frac{1}{\bar{q}}}\leq C.
$$
Here
$$
q>\frac{2\bar{r}(m_1\bar{r}-2)}{2^*l(\bar{r}-2)},\qquad q>\frac{2\bar{r}(m_2\bar{r}-2)}{2^*l(\bar{r}-2)}
$$
and $l=2$ or $l=m$ if $\alpha_1, \alpha_2\geq \frac{1}{2}$, $l=\max(2-N(1-2\alpha_1), 2-N(1-2\alpha_2)) $ or $l=\max(m, 2-N(1-2\alpha_1), 2-N(1-2\alpha_2))$ if $\alpha_1, \alpha_2<\frac{1}{2}$ in (\ref{4102}).


\begin{thebibliography}{20}
\bibitem{Bass}
F. G. Bass and N. N. Nasanov,  Nonlinear electromagnetic spin waves, {\it  Phys. Rep.},  189(1990), 165--223.



\bibitem{Borovskii}
A. V. Borovskii and A. L. Galkin, Dynamical modulation of an ultrashort high-intensity laser pulse in matter, {\it JETP}, 77(1993), 562--573.

\bibitem{Bouard}
A. de Bouard, N. Hayashi and J. C. Saut,  Global existence of small solutions to a relativistic nonlinear
Schr\"{o}dinger equation, {\it  Commun. Math. Phys.}, 189(1997), 73--105.



\bibitem{Cazenave}
 T. Cazenave, Semilinear Schr\"{o}odinger equations, Courant Lecture Notes in
Mathematics 10, New York University, Courant Institute of Mathematical
Sciences, AMS, Providence, RI, 2003.





\bibitem{Colin1}
M. Colin, On the local well-posedness of quasilinear Schr\"{o}dinger equations in arbitrary space dimension,
{\it Commun. Partial Diff. Eqns}, 27(2002), 325--354.

\bibitem{Ginibre1}
J. Ginibre and G. Velo, On a class of nonlinear Schr\"{o}dinger equations, {\it J. Funct. Anal.}, 32(1979), 1--71.

\bibitem{Ginibre2}
J. Ginibre and G. Velo, On a class of nonlinear Schr\"{o}dinger equations with non local interaction, {\it Math. Z.}, 170(1980), 109--136.

\bibitem{Glassey}
R. T. Glassey, On the blowing up of solutions to the Cauchy
problem for nonlinear Schr\"{o}dinger equations, {\it J. Math. Phys}.,
18(1977), 1794--1797.


\bibitem{Goldman}
M. V. Goldman and M. Porkolab,  Upper hybrid solitons ans oscillating two-stream instabilities, {\it Phys. Fluids},
19(1976), 872--881.

\bibitem{Guo}
B. L. Guo, J. Q. Chen and F. Q. Su, The ``blow up" problem for a quasilinear Schr\"{o}dinger equation, {\it J. Math. Phys.}, 46(2005), 073510, 10 pp.


\bibitem{Kenig}
C. E. Kenig, G. Ponce and L. Vega,  The Cauchy problem for quasi-linear Schr\"{o}dinger equations, {\it Invent. Math.},
158(2004), 343--388.

\bibitem{Litvak}
A. G. Litvak and A. M. Sergeev, One dimensional collapse of plasma waves,  {\it JETP Lett.}, 27(1978), 517--520.


\bibitem{Makhankov}
V. G. Makhankov and V. K. Fedynanin, Non-linear effects in quasi-one-dimensional models of condensed
matter theory, {\it  Phys. Rep.}, 104(1984), 1--86.


\bibitem{Poppenberg1}
M. Poppenberg, On the local well posedness of quasi-linear Schr\"{o}dinger equations in arbitrary space
dimension,  {\it J. Differential Equations}, 172(2001), 83--115.



\bibitem{Ritchie}
B. Ritchie,  Relativistic self-focusing and channel formation in laser-plasma interactions, {\it Phys. Rev. E},
50(1994), 687--689.


\bibitem{SW2}
X. F. Song, Z. Q. Wang, How Hartree type nonlinearity takes effect on the properties for the solution of a quasilinear Schr\"{o}dinger equation,
preprint.

\end{thebibliography}
\end{document}